\documentclass[twocolumn,twoside,slac_two]{revtex4}
\usepackage{graphicx}
\usepackage{fancyhdr}
\begin{document}

\title{Model analysis of the very high energy detections of the starburst galaxies M82 and NGC 253}

%

\author{E. de Cea del Pozo, D. F. Torres, A. Y. Rodr\'iguez}
\affiliation{Institut de Ciencies de l'Espai (IEEC, CSIC), Barcelona (Spain)}
\author{O. Reimer}
\affiliation{Institut f\"ur Astro- und Teilchenphysik, Universit\"at Innsbruck \& KIPAC, Stanford University}

\begin{abstract}
In the light of recent results detecting the nearest starburst galaxies M 82 and NGC 253 at very high energy (VHE), we present a multi-wavelength model succesfully explaining their gamma-ray diffuse emission. The detection of M 82 was presented by VERITAS collaboration in the recent ICRC, while the integral flux of the fainter NGC253 has just been published by HESS Collaboration. Also, in the 2009 Fermi Symposium itself, Fermi data coming from the direction of both galaxies was released, thus confirming their detection in the high energy (HE) regime. Slight and reasonable variations in the space parameter of already published models can fully account for the HE and VHE emission coming from M 82 and NGC 253, while agreeing with previous data detected from radio to infrared. We explore these changes and the implications they have for the cosmic-ray distribution in these galaxies.
\end{abstract}

\maketitle

\thispagestyle{fancy}


\section{Introduction}
Starburst galaxies were anticipated as $\gamma-$ray sources~\cite{paglione, persic}, provided sufficient instrumental sensitivity, due to their enhance star formation and supernova (SN) explosion rate, in dense (gas and dust enriched) environments. SN remnants and shock winds from massive stars are supposed to accelerate cosmic rays (CR). Due to their collisions with ambient nuclei and subsequent $\pi^{0}$ decay, very energetic $\gamma-$rays are produced, and those can in turn be detected both with space-born and ground-based imaging atmospheric Cherenkov telescopes.

Recently, the detection of M82 was presented by the VERITAS collaboration at the ICRC 2009, while the integral flux of the fainter NGC 253 has just been published by the H.E.S.S. collaboration. During the 2009 Fermi Symposium, the data collected by the Fermi space telescope was released and found coming in the direction of both galaxies.

Slight and reasonable variations in the parameter space of already published models~\cite{domingo, decea} can fully account for the HE and VHE emission coming from both galaxies, while agreeing with previous data detected from radio to infrared (IR). We explore these changes and some implications they have for the CR distribution in these galaxies.

\section{Description of the model}
The aim of this and previous studies is to perform a multi-wavelength model for the emission coming from the central part of the starburst galaxies M82 and NGC 253. To model the highest energy emission we use the $Q-Diffuse$ code, already presented in~\cite{torres} and improved in~\cite{domingo} and~\cite{decea}. The computation consists on solving the diffusion-loss equation for a steady state population of both electrons and protons, taking into account losses in each particle population, and also secondary production.

\begin{figure*}[t]
\centering
\includegraphics[width=135mm]{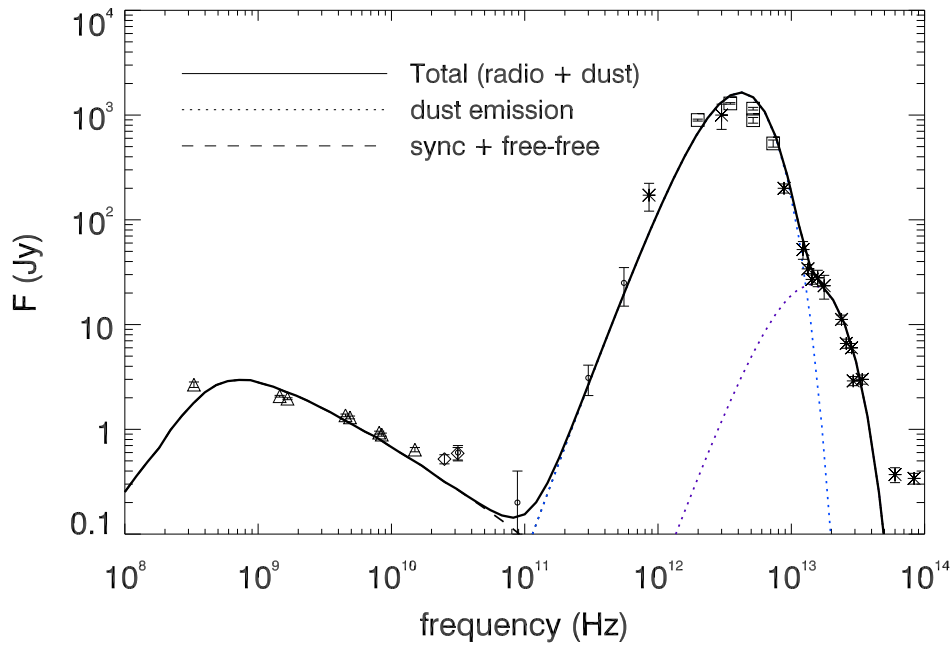}
\caption{Multifrequency spectrum of NGC 253 from radio to infrared. The observational data points correspond to: Carilli (triangles, 1996), Elias et al. (circles, 1978), Rieke et al. (asterisks, 1973), Ott et al. (diamonds, 2005) and Telesco  et al. (squares, 1980) (and references therein). The results from modelling correspond to: synchrotron plus free-free emission (dashed), dust emission (dotted) splitted in a cool (blue, $T_{cold} = 45 K$) and a warm (purple, $T_{warm} ~ 200 K$) component, and the total emission from radio and IR emission (solid).}
\label{radioIR}
\end{figure*}

One of the major achievements of this study is presenting an accurate fit to multi-wavelength data. In Fig.~\ref{radioIR}, a multi-frequency spectrum is overplotted to previous radio and IR data. Note that the prediction from the model is the summed contribution of synchrotron, free-free and dust emission, and not an actual fit to radio or IR data.

\section{Galaxy M 82}
The near, almost edge-on starburst galaxy M82 has a low-mass gas content, mostly concentrated in the inner 2 kpc, and presents a high luminosity both in the far IR and X-ray domain. As part of the M81 group, M82 shows hints of an encounter with some of its members 1Gyr ago. As a result of tidal forces, it harbors a central (300 pc) starburst. 

\begin{figure*}
\centering
\includegraphics[width=135mm]{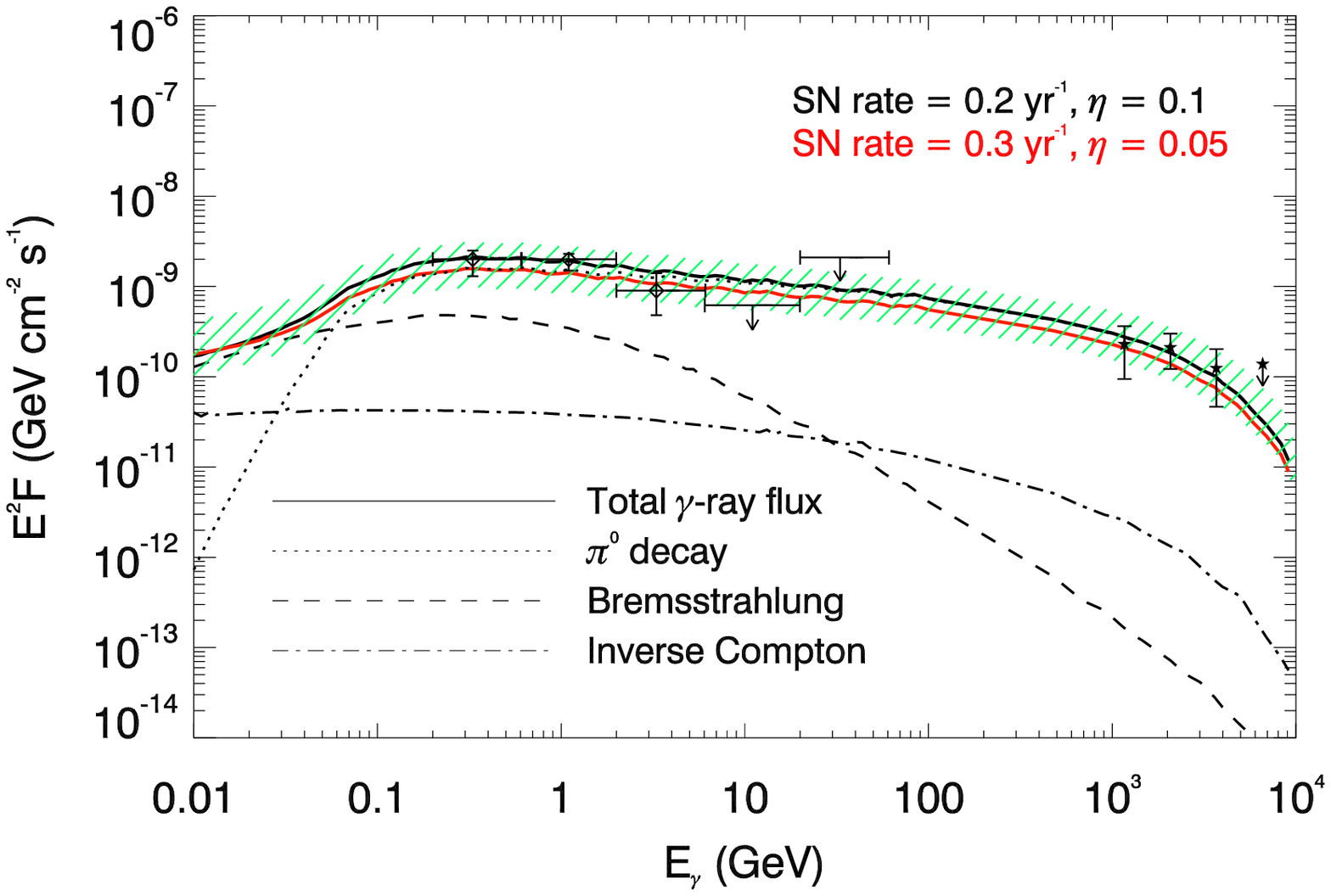}
\caption{Energy distribution of the differential gamma-ray fluxes of M82, exploring a range of uncertainties in supernova explosion rate and efficiency to inject energy from SN to CR. The shaded green area corresponds to the original model of de Cea del Pozo et al. (2009). Data points and upper limit correspond to both VERITAS (stars) and Fermi (diamonds) detections.}
\label{sedM82}
\end{figure*}

VHE $\gamma-$ray emission coming from M82 was claimed by VERITAS~\cite{veritas} during this summer, while the HE regime was covered by Fermi space telescope from the same galaxy. In Fig.~\ref{sedM82}, this data is shown together with the spectral energy distribution (SED) of the already published model~\cite{decea} for a range of parameters and some specific outputs to better predict observed results (see Table~\ref{parametM82}). A separate contribution is plotted coming from each $\gamma-$ray generator: neutral pion ($\pi^{0}$) decay, bremsstrahlung and inverse Compton radiation. The latter was computed having the cosmic microwave background (CMB), far and near IR photon densities as targets altogether (see Fig.~\ref{radioIR}). As can be seen, $\pi^{0}$ decay contribution dominates at VHE energies.

\begin{table*}
\begin{center}
\caption{Physical parameters used in the multi-wavelength model of M82, both in~\cite{decea} and the present study, in which values are specified in order to match with the VERITAS detection. The (small) variations explored here are, in any case, inside the former predictions of the original model. Numbering divide the list of parameters in: 1) observational values, 2) derived from 1), 3) obtained from modelling, and 4) assumed. SB stands for starburst.}
  \begin{tabular}{ | l | l | l | c |}
    \hline
    & \textbf{Physical parameters} & \textbf{de Cea del Pozo et al (2009)} & \textbf{VERITAS-driven model} \\ 
    \hline 
    & Distance & 3.9 $\pm0.3_{random}$ $\pm0.3_{systematic}$ Mpc & \ldots \\
    &Inclination & 77$\pm 3$ $^\circ$ & \ldots \\
    & Radius SB & 300 pc & \ldots \\
    & Radius Disk & 7 kpc & \ldots \\
    & Height SB & 200 pc & \ldots \\
    & Gas Mass SB & $2 \times 10^8 M_\odot$ ($H_2$) & \ldots \\
    & Gas Mass Disk & $7 \times 10^8 M_\odot$ (HI), & \ldots \\
    (1) & & $1.8 \times 10^9 M_\odot$ ($H_2$) & \\
    & IR Luminosity & $4 \times 10^{10} L_\odot$ & \ldots \\
    & SN explosion rate & 0.3 yr$^{-1}$ (0.1 yr$^{-1}$) & 0.2 yr$^{-1}$ | 0.3 yr$^{-1}$ \\
    & SN explosion energy & 10$^{51}$ erg & \ldots \\
    & SN energy transferred to CR & 10\% & 10\% | 5\% \\
    & Convective velocity & 600 km s$^{-1}$ & \ldots \\
    & Dust temperature & 45 K & \ldots \\
    & Ionized temperature & 10000 K & \ldots \\
    \hline
    (2) & Uniform density SB & $\sim 180 cm^{-3}$ & \ldots \\
    \hline
    & Dust emissivity index & 1.5 & \ldots \\
    & Emission measure & $5 \times 10^5$ pc cm$^{-6}$ & \ldots \\
    (3) & Magnetic field & $120$ $\mu$G  ($270$ $\mu$G) & $170$ $\mu$G | $210$ $\mu$G \\
    & Proton to electron primary ratio & 50 (30) & \ldots \\
    & Slope of primary injection spectrum & 2.1 & \ldots \\
    \hline
    & Maximum energy for primaries & $10^6$ GeV & \ldots \\
    (4) & Diffusion coeficient slope & 0.5 & \ldots \\
    & Diffusive timescale & $1-10$ Myr & \ldots \\
    \hline 
  \end{tabular}
    \label{parametM82}
\end{center}
\end{table*}

\section{Galaxy NGC 253}
The also near, barred-spiral starburst galaxy NGC 253 has been deeply studied through the years. Its continuum spectrum  peaks in the far IR with a high luminosity. Its inner (100 pc) region is characterized, as well as M82, by starburst activity. 

\begin{figure*}
\centering
\includegraphics[width=135mm]{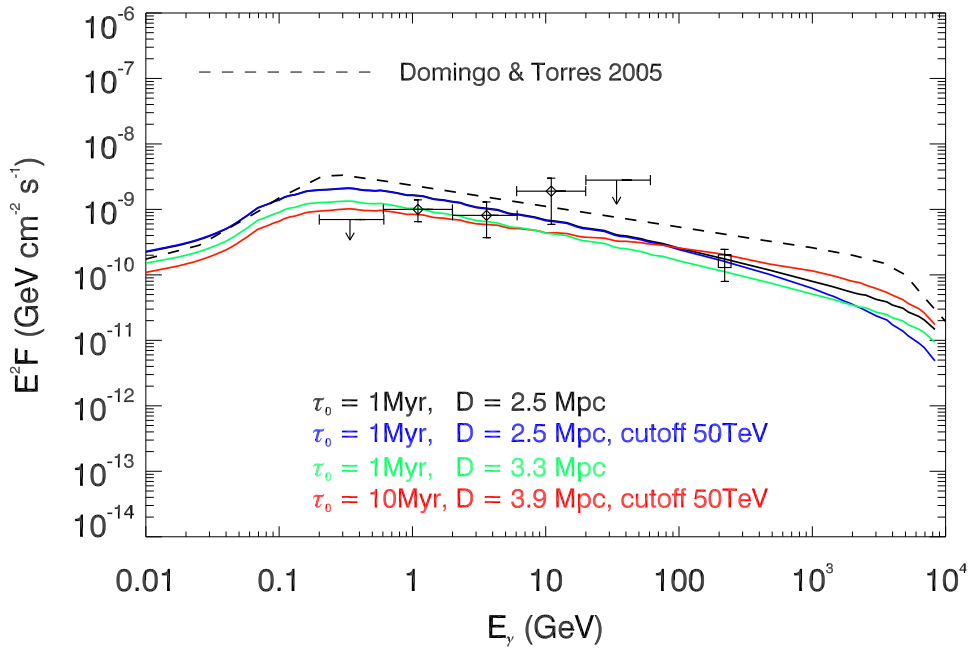}
\caption{Energy distribution of the differential gamma-ray fluxes of NGC 253, exploring the uncertainty in distance, a range of timescale diffusion ($\tau_{0}$) and possible cutoffs in the proton injection spectrum. The original model from Domingo \& Torres (2005) is also shown for comparison, as well as data points from Fermi detection (diamonds) and the integral flux from the H.E.S.S. detection (square) transformed in differential flux (assuming a range of injection spectra).}
\label{sedNGC253}
\end{figure*}

The integral flux published by H.E.S.S. constraints previously predictions for NGC 253 at the VHE regime. Moreover, the first results from the Fermi telescope add more information to the spectrum of the galaxy at lower energies. A set of curves for the SED are specifically plotted in Fig.~\ref{sedNGC253} to achieve this low flux, exploring uncertainties in the distance to this galaxy and subsequent ranges in the magnetic field and diffusive timescale. Apart from diffusing away during $10^{6 - 7}$ yrs (see Table~\ref{parametNGC}), particles can escape the inner starburst convectively, carried away by winds ($\sim 300$ km s$^{-1}$), and through pion collisions with ambient gas, in even shorter timescales of around a few $10^{5}$ yrs. Since the diffusion timescale of the particle depends on the energy, the shorter it is, the steeper the $\gamma-$ray spectrum gets (the higher the losses are). The diffusion coefficient associated is $\sim 10^{26 - 27}$ cm$^{-2}$ s$^{-1}$ at $1 - 10$ GeV, compared to the $\sim 10^{28}$ cm$^{-2}$ s$^{-1}$ value in our Galaxy.

\begin{table*}
\begin{center}
\caption{Physical parameters used in the multiwavelength model of NGC 253, as presented both in the previous~\cite{domingo} paper and in the present study, which explores some variations allowed within the model in order to explain H.E.S.S. detection. Numbering divide the list of parameters in: 1) observational values, 2) derived from 1), 3) obtained from modelling, and 4) assumed. SB stands for starburst.}
  \begin{tabular}{ | l | l | l | c |}
    \hline
    & \textbf{Physical parameters} & \textbf{Domingo \& Torres 2005} & \textbf{H.E.S.S.-driven model} \\ \hline 
    & Distance & 2.5 Mpc & 2.6 Mpc | 3.9 Mpc \\
    & Inclination & 78$^\circ$ & \ldots \\
    & Radius SB & 100 pc & \ldots \\
    & Radius Disk & 1 kpc & \ldots \\
    & Height SB & 70 pc & \ldots \\
    & Gas Mass SB & $3 \times 10^7 M_\odot$ & \ldots \\
    (1) & Gas Mass Disk & $2.5 \times 10^8 M_\odot$ & \\
    & IR Luminosity & $(2 - 4) \times 10^{10} L_\odot$ & \ldots \\
    & SN explosion rate & 0.08 yr$^{-1}$ & \ldots \\
    & SN explosion energy & 10$^{51}$ erg & \ldots \\
    & Convective velocity & 300 km s$^{-1}$ & \ldots \\
    & Dust temperature & 50K & 45 K  \\
    & Ionized temperature & 10000 K & \ldots \\
    \hline
    (2) & Uniform density SB & $\sim 600 cm^{-3}$ & \ldots \\
    \hline
    & Dust emissivity index & 1.5 & \ldots \\
    & Emission measure & $5 \times 10^5$ pc cm$^{-6}$ & \ldots \\
    (3) & Magnetic field & 300 $\mu$G & 200 $\mu$G | 270 $\mu$G \\
    & Proton to electron primary ratio & 50 & 30 \\
    & Slope of primary injection spectrum & 2.3 & 2.1 \\
    \hline
    & Maximum energy for primaries & $10^6$ GeV & \ldots \\
    (4) & Diffusion coeficient slope & 0.5 & \ldots \\
    & Diffusive timescale & 10 Myr & 1 Myr | 10 Myr  \\
    \hline 
  \end{tabular}
  \label{parametNGC}
\end{center}
\end{table*}

\section{Concluding remarks}
Our multi-wavelength model explains reasonable well both the HE and VHE emission coming from the two closest starburst galaxies M82 and NGC 253, within a range of explored parameters. Every component of the emission can be tracked to one and the same original CR population, and, in turn, result as a consequence of all electromagnetic and hadronic channels from the primary and subsequently-produced secondary particles.

CR enhancement present in these starburst galaxies is reflected in the high energy density values that can be obtained from the steady proton population. Above a proton energy of $\sim1500$ GeV (corresponding to E$_{\gamma}$ $\sim~ 250$ GeV), the energy density is around 10 eV cm$^{-3}$ for M82 and similar value for NGC 253.

Now that the VHE regime has been finally achieved by ground-based telescopes, and Fermi has proven that detections are possible from lower energies (~ 100 MeV), 
a full picture of $\gamma-$ray emission is begining to appear more and more clearly.

\bigskip 
\begin{acknowledgments}
We acknowledge support by grants AYA2006-00530 and SGR2009-811. \\
\\

The work of E. de Cea del Pozo has been made under the auspice of a FPI Fellowship, grant BES-2007-15131.
\end{acknowledgments}

\bigskip 

\end{document}